\documentclass[fleqn,usenatbib]{mnras}

\pdfoutput=1
\usepackage{hyperref} 
\usepackage[T1]{fontenc}
\usepackage{ae,aecompl}
\usepackage{slantsc}
\usepackage{float}

\usepackage{graphicx}	% Including figure files
\usepackage{amsmath}	% Advanced maths commands
\usepackage{amssymb}	% Extra maths symbols
\usepackage{subfigure}
\usepackage{etoolbox}
\usepackage{soul}
\makeatletter
\patchcmd\@combinedblfloats{\box\@outputbox}{\unvbox\@outputbox}{}{\errmessage{\noexpand patch failed}}
\makeatother

\title[The locally measured Hubble parameter]{The impact of the locally measured Hubble parameter on the mass of Sterile neutrino} % I excluded /LaTeXe/

\author[M. Ebadinejad.]{
M. Ebadinejad,$^{1}$\thanks{E-mail: m.ebadinejad@gmail.com (ME)}
\\
$^{1}$Astronomy Centre, Department of Physics and Astronomy, University of Sussex, East Sussex, Brighton BN1 9QH, UK\\
}

\date{Accepted XXX. Received YYY; in original form ZZZ}

\pubyear{2019} 

\begin{document}
\label{firstpage}
\pagerange{\pageref{firstpage}--\pageref{lastpage}}
\maketitle

\begin{abstract}
We present a precise analysis to test hypothetical models involving sterile neutrinos beyond the standard flat-$\Lambda$CDM cosmology with the CMB observations from the \emph{Planck} mission and BAO measurements. This analysis shows that adding the locally measured Hubble parameter $H_{0}$ = $73.00\pm1.75 $  km $\textrm s^{-1}$ Mp$\textrm c^{-1} $ to the data removes the need for the informative physical $m_{sterile}^{thermal}$ prior in CMB constraints of $ m_{\nu,sterile}^{eff}$. Under the constraints from the data containing the locally measured $H_{0}$ we obtain an upper limit $ m_{\nu,sterile}^{eff} < 0.306 $ eV scale mass for the massive sterile neutrino, and an upper limit $\Sigma m_{\nu} < 0.214$ eV scale mass for the three degenerate massive neutrino (95 per cent confidence level). We also obtain the value $\sigma_{8} $ = $0.81^{+0.05}_{-0.06} $ (95 per cent confidence level), which is in compatibility with the constraints from \emph{Planck} 2015 CMB data at the 1$\sigma$ level. We find that introducing parameter $ m_{\nu,sterile}^{eff}$ to the model of cosmology reduces the $\sigma_{8} $ value and moves it closer to the obtained value for this parameter from the KiDS-450 analysis. Our results show that the locally measured Hubble parameter can increase constraints on $\sigma_{8} $ values.
  
%This is a simple template for authors to write new MNRAS papers.
%The abstract should briefly describe the aims, methods, and main results of the paper.
%It should be a single paragraph not more than 250 words (200 words for Letters).
%No references should appear in the abstract.
 
\end{abstract}


%%%%%%%%%%%%%%%%%%%%%%%%%%%%%%%%%%%%%%%%%%%%%%%%%%

%%%%%%%%%%%%%%%%% BODY OF PAPER %%%%%%%%%%%%%%%%%%

\section{Introduction}\label{Int}
If they exist, Sterile neutrinos would make revolutionary changes in our understanding of physics from the smallest to the largest scales. The assumption that the sterile species are thermalized along with the three flavours of massive (active) neutrinos in the early Universe created immense motivation amongst particle physicists and cosmologists to search for this hypothetical particle. They interact only via the gravitational force, rather than the fundamental forces of the Standard Model (SM) of particle physics. Hence, their existence would introduce physics beyond the SM.
\par
In particle physics, from the Z-Boson experiment in the Large Electron-Positron Collider (LEP), the number of active (light) neutrinos which are sensitive to the weak interactions of the SM is three, corresponding to the three flavours of neutrino \citep{2004PhLB..592....1P}. However, it is also possible to have the number of mass states in neutrinos greater than the number of active flavour states. In this case, the results of the LEP experiment are justified, and the extra neutrino states must be considered as a sterile neutrino. The anomalies in Short Baseline (SBL) neutrino oscillation measurements, for instance the results of the Liquid Scintillator Neutrino Detector (LSND) experiment, provide independent evidence of neutrino conversions at a greater mass splitting difference of masses squared. In this case a fourth neutrino, a light sterile with a mass of $ \sim $(1eV), is required \citep{2001PhRvD..64k2007A, 2013MPLA...2830004P}. Also the results of the MiniBooNE experiment indicate the conversions of candidate flavour neutrinos to a sterile at SBLs \citep{2007PhRvL..98w1801A}. This experiment in Fermilab reports the results from an analysis of electron-neutrino ($\nu_{e}$) appearance data in SBLs, assuming a mass oscillation which corresponds to a sterile neutrino \citep{2018PhRvL.121v1801A}.
\par
Cosmology is significantly affected by sterile neutrinos from the radiation-dominated era in the early Universe to the late Universe with matter and radiation energy density \citep{2017PhR...711....1A}. The light sterile neutrinos with eV-scale mass, and the relic of massive active neutrinos, contribute to the total radiation energy density of the Universe as the characteristic dark radiation. Where the Hubble Law dominates the expansion, they fix the expansion rate in the radiation-dominated era of the Universe \citep{2006PhR...429..307L}. They affect cosmic microwave background (CMB) anisotropies \citep{2002ARA&A..40..171H, 1995ApJ...455....7M}. 
The anisotropies of the CMB contain information on the mass and energy content of the Universe and the origin of cosmic neutrinos from the early Universe \citep{2015APh....63...66A}. Hence, cosmological observations of the CMB can potentially provide us with evidence for the existence of sterile species, independent of the laboratory experiments in particle physics  \citep{2017PhR...711....1A}. CMB observations are useful to probe the predictions beyond the flat Lambda-Cold Dark Matter ($\Lambda$CDM) model of cosmology \citep{2004S&W....43b..90W, 2013neco.book.....L}. There is the assumption that sterile neutrinos are partially or completely thermalized at the decoupling time in the early Universe and they affect the CMB \citep{2015APh....63...66A, 2016arXiv160101475G}.
The constraints on the linear matter power spectrum of thermalized sterile neutrinos at eV-scale mass and the constraints on massive active neutrinos are similar as both are measured in large scale structure (LSS) observations when they are non-relativistic \citep{2017PhR...711....1A}. In a similar way to active neutrinos, the eV-scale mass thermalized sterile neutrinos in the Big Bang nucleosynthesis (BBN) epoch contribute to the total radiation energy density when they are relativistic. They significantly alter the effective number of relativistic species ($N_\textrm{eff}$) in the early Universe. This is while massive neutrinos later influence matter density. Hence, BBN and LSS can constrain a partially or fully thermalized sterile neutrino \citep{1977PhLB...66..202S, 2017IJMPE..2641007S, 2018EPJWC.18401011M}.
The $N_\textrm{eff}$ is a quantity that precisely describes relativistic species, such as massive active neutrinos as well as eV-scale mass light sterile neutrinos at the BBN epoch with high relativistic energy density; when the abundant light species constrain the thermalized sterile neutrinos in the early Universe. But $N_\textrm{eff}$ does not accurately describe the non-relativistic sterile neutrinos over the period of structure formation \citep{2016RvMP...88a5004C}. In the case of KeV-scale massive sterile neutrinos-as a possible dark matter particle candidates-they do not affect BBN due to their abundance, which is very small in relative density, compared to radiation in the radiation-dominated epoch of BBN. Therefore, in this work we show that data from observation of the CMB, will only provide constraints on light sterile neutrinos. However, we also assume the
presence of massive sterile neutrinos, via partial or complete thermalization in the early Universe \citep{1998PhLB..427...97B}. The presence of a massive sterile neutrino in cosmological models provides a potential solution to the neutrino oscillation
anomalies \citep{2013JHEP...05..050K}. This follows recent research which is motivated by models including massive sterile neutrinos \citep{2013JCAP...10..044H, 2014PhRvL.112e1303B, 2014PhRvL.112e1302W}. Apart from ordinary massive neutrinos, the presence of additional massive particles, such as massive sterile neutrinos, would resolve the tension between \emph{Planck} CMB observations and other astrophysical data by leading to a lower value in the amplitude of mass fluctuation ($\sigma_{8} $), \citep{2016A&A...594A..13P}. In this paper, we investigate the constraints on massive sterile neutrinos by the \emph{Planck} observations and other data in an extended model of standard $\Lambda$CDM. We also look to the constraints on $\sigma_{8} $ and discuss whether a possible massive sterile neutrino would significantly alter this parameter.
\par 
The existence of a sterile neutrino would alter the current favoured SM of cosmology: flat-$\Lambda$CDM. CMB observations are based on cosmological parameters which
are defined at low redshift (low z) and depend on the SM. It is useful to
measure some of these parameters locally, to determine whether they are independent
of the cosmological model. The measurement of quantities such as the Hubble constant
in the local Universe helps to test the consistency of the current flat-$\Lambda$CDM
model and its constraints on cosmology, as the observations are independent of the
model itself \citep{2013PDU.....2...65V}. Hence, addition of the directly measured
Hubble parameter at low redshift to CMB cosmological data at high redshift helps one
to investigate whether the extension of the SM of cosmology by a new
parameter is validated. The proposed way to do the test is by extending the main
$\Lambda$CDM model using one or more extra parameters, and constraining these
parameters with current observations.
  
\par 
The $\Lambda$CDM model successfully fits various observations.  
 In this paper, we use the recent locally measured Hubble parameter in addition to the \emph{Planck} CMB observations and galaxy BAO data, to test the impact from the introducing sterile and massive neutrinos parameters to the standard $\Lambda$CDM, on the late cosmological parameters, $H_{0}$, $\sigma_{8} $, $\Omega_{m}$. We focus on testing the impact on the extended parameters $ m_{\nu,sterile}^{eff}$ massive sterile neutrino, $ \Sigma m _{\nu} $ three degenerate massive neutrino, and the parameter $\sigma_{8} $, from the locally measured $H_{0}$; this is what we usually can not do in standard $\Lambda$CDM due to its tension with the local Universe.

%This is a simple template for authors to write new MNRAS papers.
%See \texttt{mnras\_sample.tex} for a more complex example, and \texttt{mnras\_guide.tex}
%for a full user guide.

%All papers should start with an Introduction section, which sets the work
%in context, cites relevant earlier studies in the field by \citet{Others2013},
%and describes the problem the authors aim to solve \citep[e.g.][]{Author2012}.
\section{Methods}

\subsection{Observational data}\label{Obs}
In this work the latest observed Cosmic Microwave Background, CMB anisotropy-temperature data by the \emph{Planck} mission in 2015 \citep{2016A&A...594A..13P} is used. We consider the likelihood at $ 0 \leq l \leq 2500 $ in temperature (TT): LowTEB, $ 2 < l < 30 $ which is the low-\emph{l} part of the TT power spectrum with the range including polarization, and Plik, the CMB TT likelihood in the range of $ 30 < l < 2500 $. Also the Baryonic Acoustic Oscillation (BAO) data from the Baryon Oscillation Spectroscopic Survey (BOSS) that provides us with more information on the low redshift (Low Z) Universe \citep{2005ApJ...633..560E} is used. This is contained in four different samples: LOWZ (low redshift) and CMASS \citep{2016MNRAS.457.1770C} from updated DR12 which is the final data release of SDSS-III (Sloan Digital Sky Survey), 6dFGS (six degree field galaxy survey), \citep{2011MNRAS.416.3017B} and MGS (main galaxy sample), \citep{2015MNRAS.449..835R}. In addition, we use the other astrophysical observable which is the recent directly measured local value of the Hubble constant, $H_{0}$ = $73.00\pm1.75 $  km $\textrm s^{-1}$ Mp$\textrm c^{-1} $, by \cite{2016ApJ...826...56R} in order to find out the impact of this observable on sterile neutrinos in the analysis. Thus, we have two sets of observational data in this paper: CMB + BAO and CMB + BAO + $H_{0}$. 

\subsection{Modelling and analysis}

 We use the software called \textsc{camb}, Code for Anisotropies in the Microwave Background by \cite{2013PhRvD..87j3529L}. This is a code for cosmological calculations as well as calculating CMB and matter power spectra. In the process of modelling and analysis, the constraints on the cosmological parameters based on the $ \Lambda$CDM model are produced using \textsc{camb} Boltzmann code \citep{2000ApJ...538..473L}. These parameters are: $ \omega_{b}$ = $ \Omega_{b}h^{2}$, $ \omega_{c}$ = $ \Omega_{c}h^{2}$, 100$ \Theta_{MC}$, $\tau $, $ A_{s}$ and $n_{s}$, which are, respectively, baryon matter density, present cold dark matter density, the ratio between the acoustic horizon and the angular diameter distance at the time of decoupling, Thomson scattering optical depth, amplitude of first curvature perturbation power spectrum and the spectral index of the first curvature perturbation. However, these constraints are not reported in this work. But the constraints on the derived parameters of $\Omega_{m}$, $\sigma_{8}$ and $H_{0}$; the total matter density, the mass fluctuation amplitude, and the Hubble parameter, respectively, are reported. In addition, all the temperature nuisance parameters which are used by \emph{Planck} likelihood, are allowed to vary in this analysis. We adopt the same priors from the baseline \emph{Planck} 2015 analysis \citep{2016A&A...594A..13P, 2016A&A...594A..11P} for all the cosmological and the nuisance parameters in this paper. We use \textsc{CosmoMC} software, which is a Markov-Chain Monte-Carlo (MCMC) engine for exploring cosmological parameter space for plotting and presenting the results \citep{2002PhRvD..66j3511L}.
\par
We compare the sets of data with the theoretical models to test our speculation on the models in the $\Lambda$CDM paradigm of cosmology.
In the modelling we consider the light and massive sterile neutrinos as two cases to be added into the cosmological models; the parameter $ N_\textrm{eff}$ for the light sterile neutrino as it contributes to the dark radiation, and the additional parameter, $m_{\nu,sterile}^{eff} $, for the massive sterile neutrino. Under the assumption that the sterile neutrinos are thermalized along with the active neutrinos in the early Universe, the physical mass of sterile neutrino, $m_{sterile}^{thermal}$, which is thermally distributed, is measured via the equation below \citep{2016A&A...594A..13P}.

\par
\begin{equation} 
\label{eq:1}
m_{sterile}^{thermal}  = (N_\textrm{eff} - 3.046 )^{-3/4}  m_{\nu,sterile}^{eff}, 
\end{equation}

Where in a universe with an extra relativistic particle such as a possible sterile neutrino, we must consider the value greater than 3.046 for $N_\textrm{eff}$ avoiding a negative value for $m_{sterile}^{thermal} $.  
We use the arranged data sets from the previous subsection to constrain the parameters specified by light and massive sterile neutrinos, in the extended models. Thus, we compare the models $\Lambda$CDM, $\Lambda$CDM + $N_\textrm{eff}$ and $\Lambda$CDM + $N_\textrm{eff}$ +  $m_{\nu,sterile}^{eff} $ under the same data sets.

\section{Results and discussion}
We discuss the implications of the fitting results for sterile neutrinos, by looking to the constraints from our data sets on the extended parameters, $ N_\textrm{eff}$ and $m_{\nu,sterile}^{eff} $ as well as, the constraints on the derived parameters $\sigma_{8}$, $\Omega_{m}$ and $H_{0}$. The impact of the additional data, the locally measured $H_{0}$, on the parameters, is investigated especially. We include the parameter specified for the massive active neutrino in some of the models in order to check the constraint from the data on the massive neutrino, showing whether the models associated with this parameter could better fit the data. We keep the total mass of the three degenerate massive neutrinos $\Sigma m_{\nu}= 0.06 $ eV scale mass under the assumption of the minimal-mass normal hierarchy.

 \begin{table*}
\footnotesize
\caption{Fitting results for the cosmological models including a massive sterile neutrino $m_{\nu,sterile}^{eff}$, light sterile neutrino $N_\textrm{eff}$ and massive active neutrino $\Sigma m_{\nu}$ with $ 1\pm\sigma $ errors. We quote the 95 per cent confidence level upper limits for the parameters that cannot be well constrained.}   
\scriptsize
\centering 
\tabcolsep=0.33cm
\begin{tabular}{c c c c c c c c c} 
\hline\hline 
Model & Data&$ H_{0} $&$N_\textrm{eff}$&$m_{\nu,sterile}^{eff} $  &$\Sigma m_{\nu}$ &$ \Omega_{m} $ &$  \sigma_{8} $& Total $ \chi^{2} $\\ [0.5ex] 

\hline 
$\Lambda$CDM & CMB &67.33 $ \pm $ 0.98  & 3.046 & -- &0.06 & 0.314 $\pm$ 0.013 & 0.829 $\pm$ 0.014 & 11261.90\\
\hline

$\Lambda$CDM+Neff & CMB &$68.0^{+2.6}_{-2.9}$&$3.13^{+0.29}_{-0.33} $ &--&0.06& 0.311 $ \pm $ 0.020 & $0.834^{+0.021}_{-0.024}$ & 11261.92\\

\hline 

$\Lambda$CDM & CMB+BAO &68.07 $\pm $ 0.55 &3.046 &--&0.06&0.304 $\pm $ 0.007 & 0.829 $\pm$ 0.014& 11269.90 \\
\hline

$\Lambda$CDM  & \shortstack{CMB+BAO\\H0} &68.51 $\pm$ 0.53 &3.046 &--&0.06&0.298 $\pm $ 0.006  & 0.828 $\pm$ 0.015  & 11277.00\\
\hline
$\Lambda$CDM+$N_\textrm{eff}$ & CMB+BAO & 69.1 $ \pm $ 1.5 &3.22 $\pm$ 0.23 &-- &0.06 &0.301 $\pm$ 0.008 & 0.838 $\pm$ 0.019 & 11269.60 \\
\hline 
$\Lambda$CDM+$N_\textrm{eff}$ & \shortstack{CMB+BAO\\H0}  &70.7$ \pm $1.2& 3.45 $ \pm $ 0.20 &-- & 0.06 &0.294 $\pm$ 0.007 & 0.851 $\pm$ 0.019 &11273.20  \\
\hline
\shortstack{$\Lambda$CDM+ $N_\textrm{eff}$ \\ $m_{\nu,sterile}^{eff}$}& CMB+BAO & $69.38^{+0.94}_{-1.6}$& $ < 3.73 $ &$ < 0.468 $ &0.06 & 0.303 $ \pm $ 0.008 & $0.821^{+0.027}_{-0.024}$ & 11269.50  \\ 
\hline
\shortstack{$\Lambda$CDM+ $N_\textrm{eff}$ \\$m_{\nu,sterile}^{eff}$}& \shortstack{CMB+BAO\\H0} &70.8 $\pm$ 1.2&3.54 $\pm$ 0.22 &$ < 0.306 $ &0.06 &0.296 $\pm$ 0.007 & $0.835^{+0.027}_{-0.023} $ & 11272.70    \\
\hline
\shortstack{$\Lambda$CDM+ $N_\textrm{eff}$ \\ $\Sigma m_{\nu} $}& CMB+BAO &69.1 $\pm$ 1.5 & 3.24 $\pm$ 0.25 &-- &$<0.240$ & 0.302 $ \pm $ 0.008 &0.834 $ \pm $ 0.022 & 11269.34\\
\hline

\shortstack{$\Lambda$CDM+$N_\textrm{eff}$ \\ $ \Sigma m_{\nu} $}&\shortstack{CMB+BAO\\$H_{0}$} & $70.6^{+1.4}_{-1.2} $& $3.44^{+0.25}_{-0.20} $ &-- &$<0.214$ & 0.295 $\pm $ 0.007 & 0.847 $\pm$ 0.021 & 11272.49\\
\hline

\shortstack{$\Lambda$CDM+ $N_\textrm{eff}$ \\ $m_{\nu,sterile}^{eff}$ \\$\Sigma m_{\nu} $}& CMB+BAO &$69.38^{+0.99}_{-1.6}$ &$ <3.73 $ &$ < 0.476 $ &$<0.216 $ &0.303 $ \pm $ 0.008  & $0.819^{+0.030}_{-0.025} $& 11269.46\\

\hline 
\shortstack {$\Lambda$CDM + $N_\textrm{eff}$ \\ $m_{\nu,sterile}^{eff}$ \\ $ \Sigma m_{\nu} $}& \shortstack{CMB+BAO \\ $ H_{0} $} &$70.7^{+1.3}_{-1.2} $ & $ 3.54^{+0.26}_{-0.22}$ &$ < 0.320 $ & $ <0.213 $ &0.297 $\pm $ 0.007 &$0.832^{+0.025}_{-0.022} $ & 11272.90\\
\hline

\end{tabular}
\label{Table:1}
\end{table*}

\subsection{Light Sterile Neutrinos}

As explained in Section \ref{Int}, light sterile neutrinos serve as candidates for the dark radiation. $N_\textrm{eff}$ = 3 is for the three SM neutrinos that were thermalized and decoupled well before the electron-positron annihilation in the early Universe. In standard cosmology it is predicted that $N_\textrm{eff}$ = 3.046 for the three active neutrinos. This is because of the transfer of a small amount of entropy due to the non-instantaneous decoupling of neutrinos at electron-positron annihilation \citep{2005NuPhB.729..221M}. In the search for a relic of an extra relativistic species, for example a light sterile neutrino, we presume that an additional light particle is produced before re-combination, with an energy
density that scales with the expansion exactly like that of active neutrinos. Here we look for a larger value for this parameter in our fitting results that may correspond to the presence of an extra light species. Hence, a measurement of extra relativistic degrees of freedom in the early Universe, $ \Delta N_\textrm{eff} $ = $N_\textrm{eff}$ - 3.046 $ > $ 0 is significant and it may suggest the existence of a light sterile neutrino or other kind of light relic particle \citep{2016A&A...594A..13P}.
\par
It is well established that allowing the parameter $N_\textrm{eff}$ to vary increases the uncertainty in the $ H_{0} $ constraint from \emph{Planck} CMB observations, alleviating the tension with the local measurement of $H_{0}$ \citep{2016JCAP...10..019B, 2016A&A...594A..13P, 2016ApJ...826...56R}. 
In this work, Table \ref{Table:1} shows that, under the constraints from CMB + BAO + $ H_{0} $ data, for the $\Lambda$CDM and the $\Lambda$CDM + $N_\textrm{eff}$ models we obtain the values of $ H_{0} $ = 68.51$ \pm $ 0.53 km $\textrm s^{-1}$ Mp$\textrm c^{-1} $, and $ H_{0} $ = 70.7 $ \pm $ 1.2 km $\textrm s^{-1}$ Mp$\textrm c^{-1} $, respectively. Here, the derived value of $ H_{0} $ for the $\Lambda$CDM model might not be reliable due to the Hubble tension which is caused by a tension metric for locally measured value of $H_{0}$ against CMB + BAO data. The difference between value of local $H_{0}$ = $73.00\pm1.75 $  km $\textrm s^{-1}$ Mp$\textrm c^{-1} $ and the derived value of $ H_{0} $ for the $\Lambda$CDM + $N_\textrm{eff}$ model, is 1.08$ \sigma $. It indicates that here the addition of the parameter $N_\textrm{eff}$ can relieve the tension between this data and the local $ H_{0} $, showing a compatibility with the local measurement; hence an extra relativistic species such as a possible light sterile neutrino in our model of cosmology, can surely improve cosmological fit to the data, as is indicated by the results of total $ \chi^{2}$ in Table \ref{Table:1}. This result is consistent with the obtained results of the similar works, for example \citep{2016A&A...594A..13P}. We also obtain a discrepancy at 2.82 $ \sigma $ level from the current \emph{Planck} 2015 CMB cosmological observations with the locally measured $H_{0}$, in the standard $\Lambda$CDM. One thing we perceive here is that the favoured flat-$\Lambda$CDM model of cosmology for large scales, which is well described by CMB observations, does not fit both measurements well, or that the systematic errors could be the cause of the tension between the local and the large scale measurements \citep{2013PDU.....2...65V, 2016JCAP...10..019B}. The fitting results also show that, the addition of BAO to CMB data leads to only a small reduction in the value of total $ \chi^{2}$ for the models, as BAO contains only a small part of the whole observational data we use and it contributes less than the CMB in fitting the models to the data combinations; but the results in Table \ref{Table:1} also show that the BAO data helps to give a better constraint on the cosmological parameters, as it produces smaller errors associated with the obtained results.

\begin{figure}
  \begin{center}
\subfigure[ ]{\includegraphics[width= 5.15cm,angle=360]{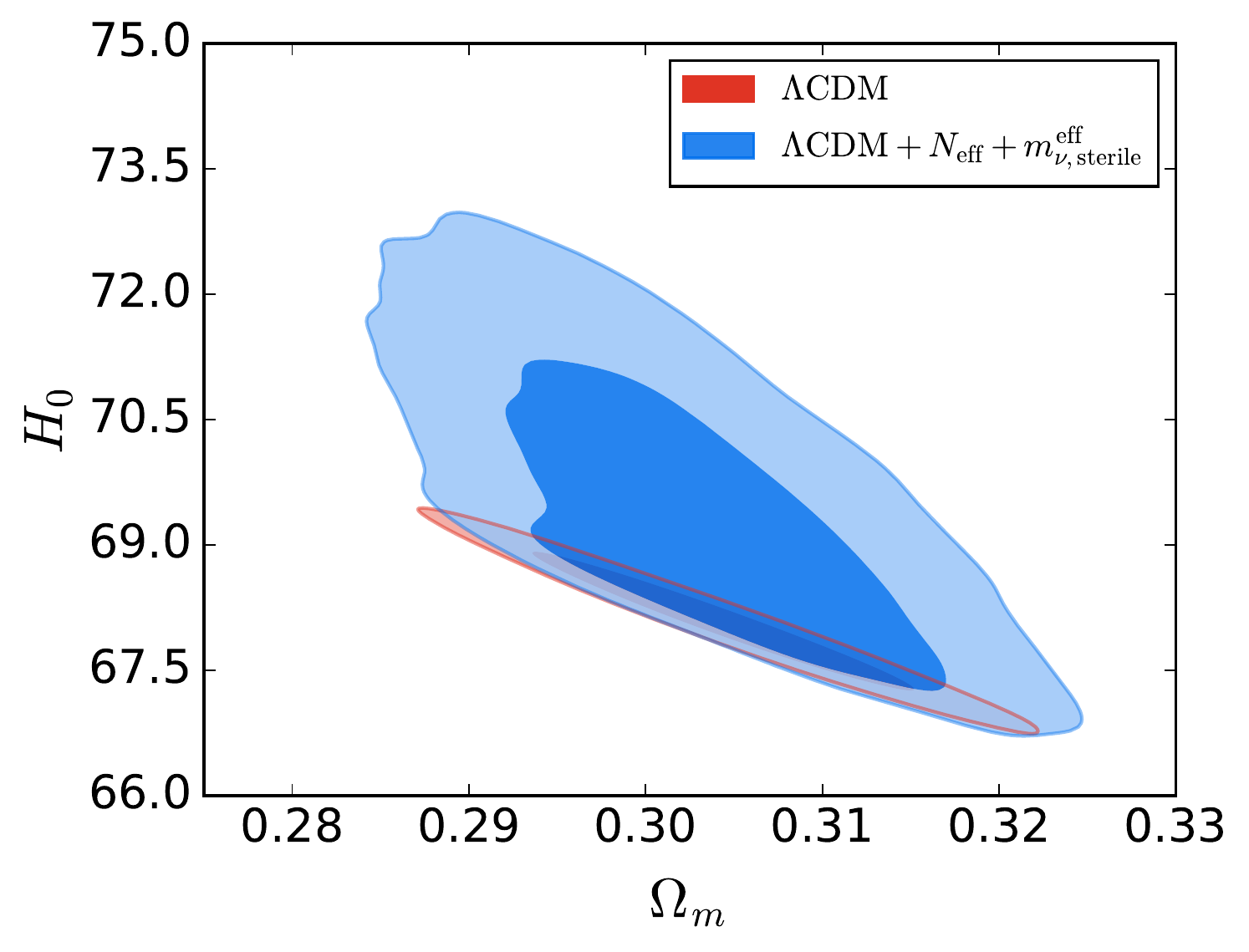}}
\subfigure[ ]{\includegraphics[width= 5.15cm,angle=360]{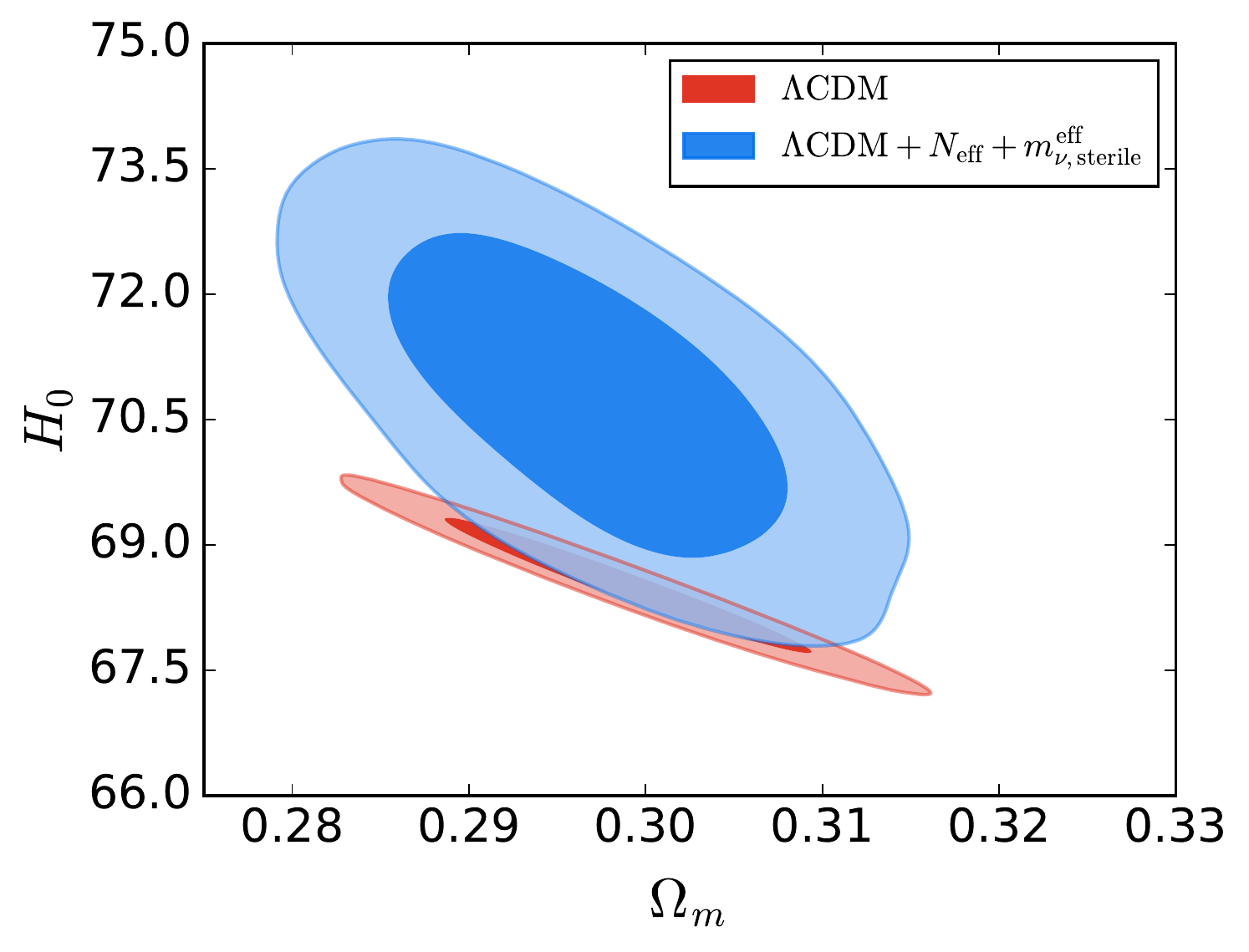}}
\caption{{The two-dimensional marginalized contours (68 and 95 per cent confidence level) in the $\Omega_{m}-H_{0}$ plane, from the constraints of the CMB + BAO (a) and the CMB + BAO + $H_{0}$ (b) data combinations in the $\Lambda$CDM and $\Lambda$CDM + $N_\textrm{eff}$ + $m_{\nu,sterile}^{eff} $ models.}}
\label{fig:1} 
\end{center} 
 
\end{figure}

The constraints from CMB + BAO were released in \emph{Planck} papers. In this work we use the updated BAO in the data set.

\subsubsection{Dark radiation constraints}

The obtained values of constraints from the data combinations on the parameter $N_\textrm{eff}$, infer the values of the extra relativistic degrees of freedom in the early Universe, given an extended $\Lambda$CDM model. In the same way as baseline \emph{Planck} 2015, we consider a wide flat prior, where we input the range (1 6) as the prior for this parameter in the analysis. Table \ref{Table:1} presents the measured value of 3.22 $ \pm 0.23 $ at the 68 per cent confidence level. Here, the CMB + BAO constraint on $N_\textrm{eff}$ provides a deviation from the standard 3.046 value only at the 0.75$\sigma$ level; as we discussed earlier, the addition of the BAO data to CMB improves the fit of the data to the models only with a small amount. Moreover, the derived value of Hubble parameter $ H_{0} $ = $69.1\pm1.5 $  km $\textrm s^{-1}$ Mp$\textrm c^{-1} $ is obtained under this data set for $\Lambda$CDM + $N_\textrm{eff}$, which is in discrepancy (below 1$\sigma$ error) with the derived $ H_{0} $ = $68.3\pm1.5 $  km $\textrm s^{-1}$ Mp$\textrm c^{-1} $ of the same analysis in the published \emph{Planck} 2015 results. This is due to the impact of the updated galaxy BAO data we use in this paper. Likewise, the CMB + BAO + $ H_{0} $ data set well constrains $N_\textrm{eff}$ in the model, with the measured value of 3.45 $ \pm 0.20 $ at the 68 per cent confidence level, providing a stronger preference at the 2.02$\sigma$ level for the extra dark radiation density. This might be an indication of the presence of extra relativistic species such as a potential light sterile neutrino. Here, the impact of the directly measured local $H_{0}$ on fitting the extended model to this data set is significant, improving the fit by $ \Delta \chi^{2} $ = 3.80 for the model.

\subsection{Massive Sterile Neutrinos} 

Here we have the parameter $m_{\nu,sterile}^{eff} $ which denotes the massive sterile
neutrino, in addition to $N_\textrm{eff}$, in the $\Lambda$CDM model. Fig.
\ref{fig:1} shows how $\Omega_{m} $ affects the
constraints on the $ H_{0} $, having a negative correlation for both
models under the data combinations. Panel(b) in the figure shows that combining the
local $ H_{0} $ measurement with the data results in the reduction of $\Omega_{m} $
values and an increase in $ H_{0} $ values for both models in comparison to the plot
in panel (a).
Table \ref{Table:1} shows that the additional $m_{\nu,sterile}^{eff}$, improves the
fit of the model to the data by only a small amount, as we find a small reduction on
the total $ \chi^{2} $ value for the model associated with this parameter in comparison to
the model with only the parameter $N_\textrm{eff}$. Under the constraint from both of
the data sets we use in this analysis, for the model including
$m_{\nu,sterile}^{eff}$, we find slightly higher derived values for parameters $
H_{0} $ and $\Omega_{m} $, and lower values for $\sigma_{8}$, in comparison with the
model containing only $N_\textrm{eff}$. It is notable that the measured values
for parameters $H_{0}$ and $\sigma_{8}$ under CMB + BAO data are in slight
disagreement with values obtained from the same analysis in the \emph{Planck} 2015
paper, due to the use of updated BAO data. Looking at these results, it
is clear that the impact of local $H_{0}$ is to increase the measured values of
$\sigma_{8}$ in the given extended SM, where in $\Lambda$CDM +
$N_\textrm{eff}$ + $m_{\nu,sterile}^{eff} $ the massive sterile neutrino helps to
pull down the $\sigma_{8}$ values but increases the derived $ H_{0} $ values. As
discussed in the baseline \emph{Planck} 2015 analysis \citep{2016A&A...594A..13P},
these results also would imply that the presence of a massive sterile neutrino in the
model of cosmology, in addition to the three degenerate massive active neutrinos
$\Sigma m_{\nu}$ as in the base $\Lambda$CDM, could potentially help to resolve the
tension between \emph{Planck} observations and the late astrophysical measurements by
introducing sufficient freedom to allow higher values of $H_{0}$ and lower values of
$\sigma_{8}$.

\subsubsection{Sterile neutrino mass constraints}

Here we consider the case of the extended model which is described earlier in this subsection. The potential degenerate mass above the minimal baseline, as well as the relativistic particle contribution from massive active
neutrinos to $N_\textrm{eff}$, is considered. Hence, we simultaneously constrain both
parameters in the extended model. For thermally distributed sterile neutrinos,
$m_{\nu,sterile}^{eff} $ is related to the physical mass via Eq.(\ref{eq:1}). To
measure the constraints on $m_{\nu,sterile}^{eff} $, we use the same prior,
$m_{sterile}^{thermal}$< 10 eV, on the physical thermal mass as the one used in the
baseline \emph{Planck} 2015 analysis; we adopt a prior with the range (3.046 10) on
parameter $N_\textrm{eff}$ to avoid negative values for the physical mass
$m_{sterile}^{thermal}$, as explained for Eq.(\ref{eq:1}).

\par 
Table \ref{Table:1} shows that, under the constraints of the CMB + BAO data for
massive sterile neutrinos, we obtain only an upper limit of $m_{\nu,sterile}^{eff} <
0.468$ eV scale mass at the 95 per cent confidence level, which is tighter than the
obtained value $m_{\nu,sterile}^{eff} < 0.571$ eV scale mass in the \emph{Planck} 2015
paper for the same analysis, due to the updated BAO data we use in this paper; under
the CMB + BAO + $ H_{0} $ data we still obtain an upper limit of $
m_{\nu,sterile}^{eff} < 0.306 $ eV scale mass at the 95 per cent confidence level,
indicating that the locally measured $ H_{0} $ only helps to constrain the mass by
tightening the constraints on the parameter $m_{\nu,sterile}^{eff} $. However, these
results are similar to the results of the baseline \emph{Planck} 2015 paper,
indicating that the current observations cannot well constrain the mass of sterile
neutrinos.
\par 
The KeV-scale massive sterile neutrinos as possible dark matter could be detectable
via X-ray observations and their effects on structure formation, if their production
originally takes place via oscillation \citep{2017PhR...711....1A}. There are also
searches for the signature of decay of this class of sterile neutrino in X-ray
observations from different sources. The best current constraint is reported from the
analysis of \emph{Chandra} X-ray observations of the Andromeda galaxy by
\cite{2014PhRvD..89b5017H}.

\subsection{Active neutrino masses}

It is found that neutrinos acquire mass via neutrino mass oscillation \citep{1998PhRvL..81.1562F}. A normal mass hierarchy with $\Sigma m_{\nu} \approx $ 0.06 eV is supposed in $\Lambda$CDM model via \emph{Planck} observations. There is also possibility of a degenerate hierarchy with larger mass for active neutrinos $\Sigma m_{\nu} \gtrsim $ 0.1 eV \citep{2016A&A...594A..13P}. This motivated us to extend our model to include massive neutrinos. A larger possible mass for neutrinos could lead to a lower $\sigma_{8}$. Neutrino masses below 1eV have only a mild effect on the CMB power spectrum as they are relativistic. But there is some sensitivity of CMB anisotropy to neutrino masses when neutrinos begin to be less relativistic at the time of recombination \citep{2016A&A...594A..13P}.   
The total $ \chi^{2}$ test results presented in Table \ref{Table:1} show that the addition of parameter $\Sigma m_{\nu}$ to the extended models, improves the fit of the model by only a small amount for both data sets.

\par
The neutrino mass constraints from other works include the analysis of the KIDS-450 data \citep{2017MNRAS.465.1454H}, a derived neutrino mass constraint leading to $\Sigma m_{\nu} < 4.0$ eV scale mass at 95 per cent confidence level is obtained by \cite{2017MNRAS.471.1259J}. In this finding \cite{2017MNRAS.471.1259J} use the implementation of massive neutrinos from \cite{2016MNRAS.459.1468M}. In this paper we use the implementation of massive neutrinos from \cite{2011JCAP...09..032L} and \cite{2012MNRAS.420.2551B}. Moreover, combination of \emph{Planck} CMB results, BAO measurements and measured Ly$\alpha$ power spectrum produces an upper limit of $\Sigma m_{\nu} < 0.14$ scale mass at 95 per cent confidence level \cite{2015JCAP...11..011P}.

\par
We use the prior with the range (0.01 1) to constrain the parameter $\Sigma m_{\nu}$. Table \ref{Table:1} shows that under the constraints of CMB + BAO data we derive an upper limit $\Sigma m_{\nu} < 0.240 $ eV scale mass at the 95 per cent confidence level for the total mass of active neutrinos which is tighter than the obtained value $\Sigma m_{\nu} < 0.266 $ eV scale mass of similar analysis in baseline \emph{Planck} 2015, here again due to the updated BAO we use. The local $ H_{0} $ measurement in the data set helps to constrain only a tighter upper limit $\Sigma m_{\nu} < 0.214$ eV scale mass at the 95 per cent confidence level. The obtained upper limit values of $\Sigma m_{\nu}$ for the model associated with massive sterile neutrino are also presented in Table \ref{Table:1}.

\begin{figure}
 \begin{center}
  \subfigure[ ]{\includegraphics[width= 5.1cm,angle=360]{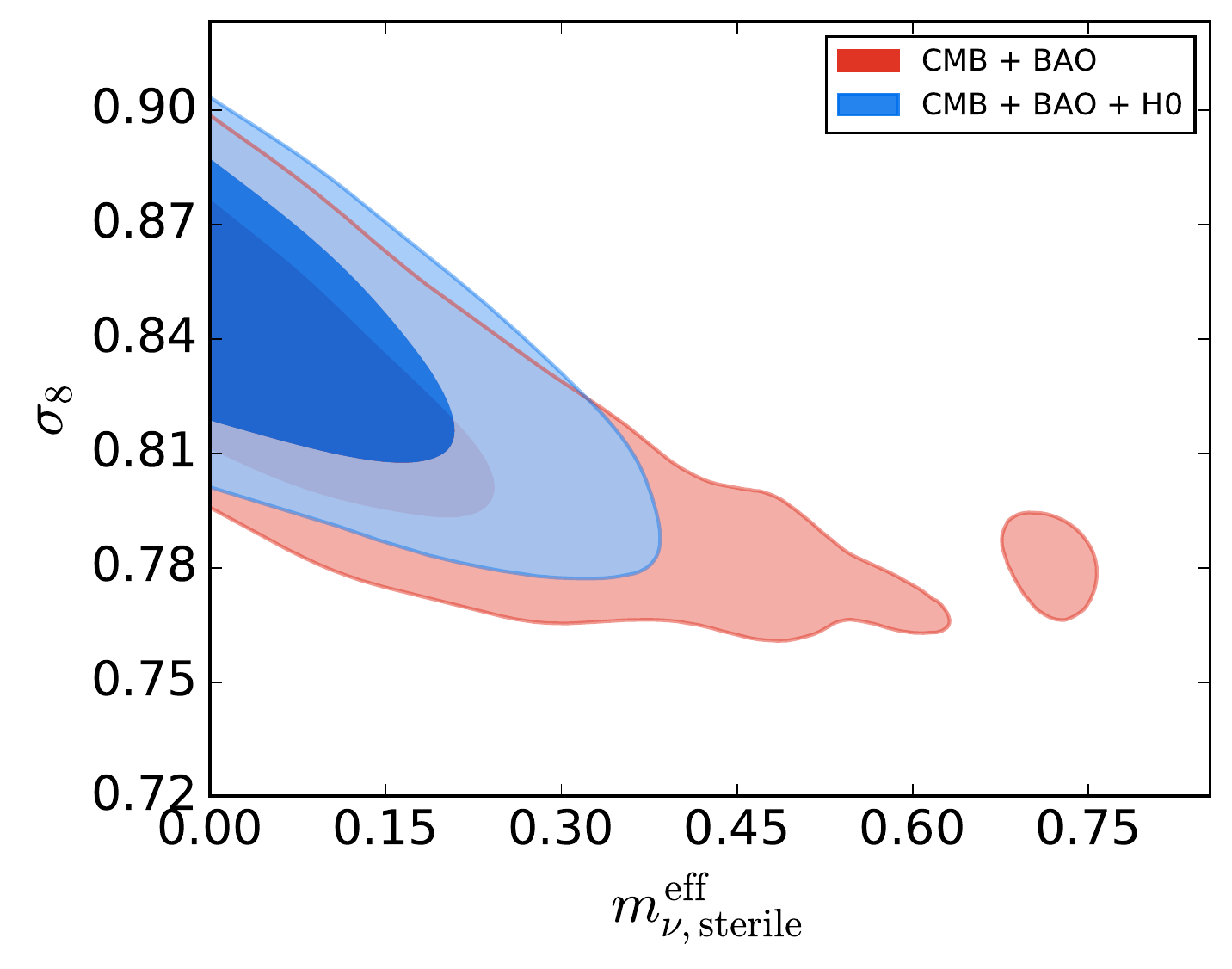}}
\subfigure[ ]{\includegraphics[width= 5.1cm,angle=360]{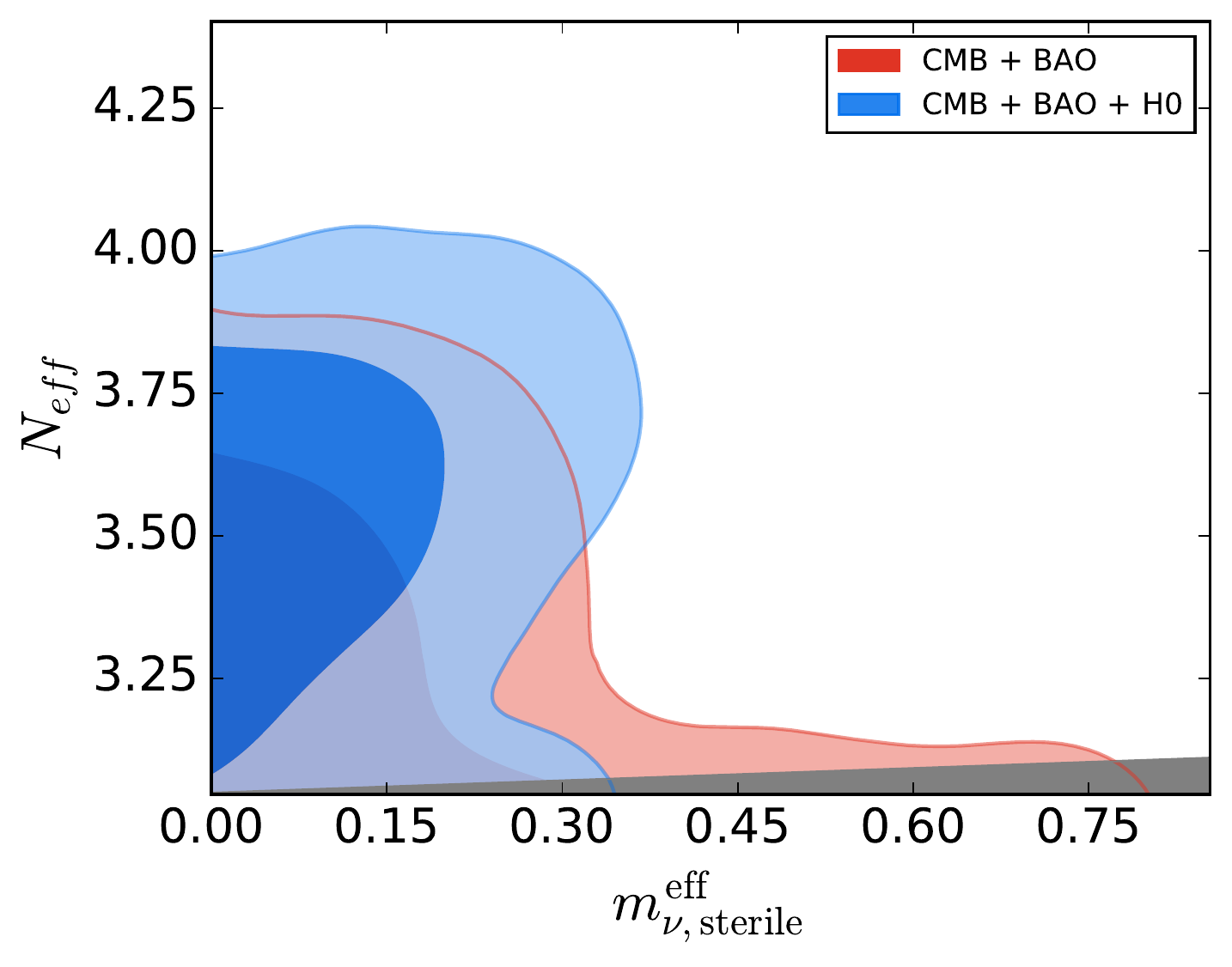}}
\caption{The constraint results for the $\Lambda$CDM + $N_\textrm{eff}$ + $m_{\nu,sterile}^{eff} $ model from the CMB + BAO and the CMB + BAO + $H_{0}$ data combinations are shown in two-dimensional marginalized posterior contours (68 and 95 per cent confidence level). Panel(a): Shows the significant impact from local $H_{0}$ on the mass of high mass sterile neutrinos. Panel(b): The grey shading shows the excluded region of parameter space by the prior $m_{sterile}^{thermal}$ < 10 eV, where the neutrinos with the high mass near to the tail of prior, behave like dark matter.} 
\label{fig:2}
\end{center}

 \end{figure}

 \subsection{Constraints on $\sigma_{8}$}

\begin{figure*}  
 
\includegraphics[width=0.65\linewidth]{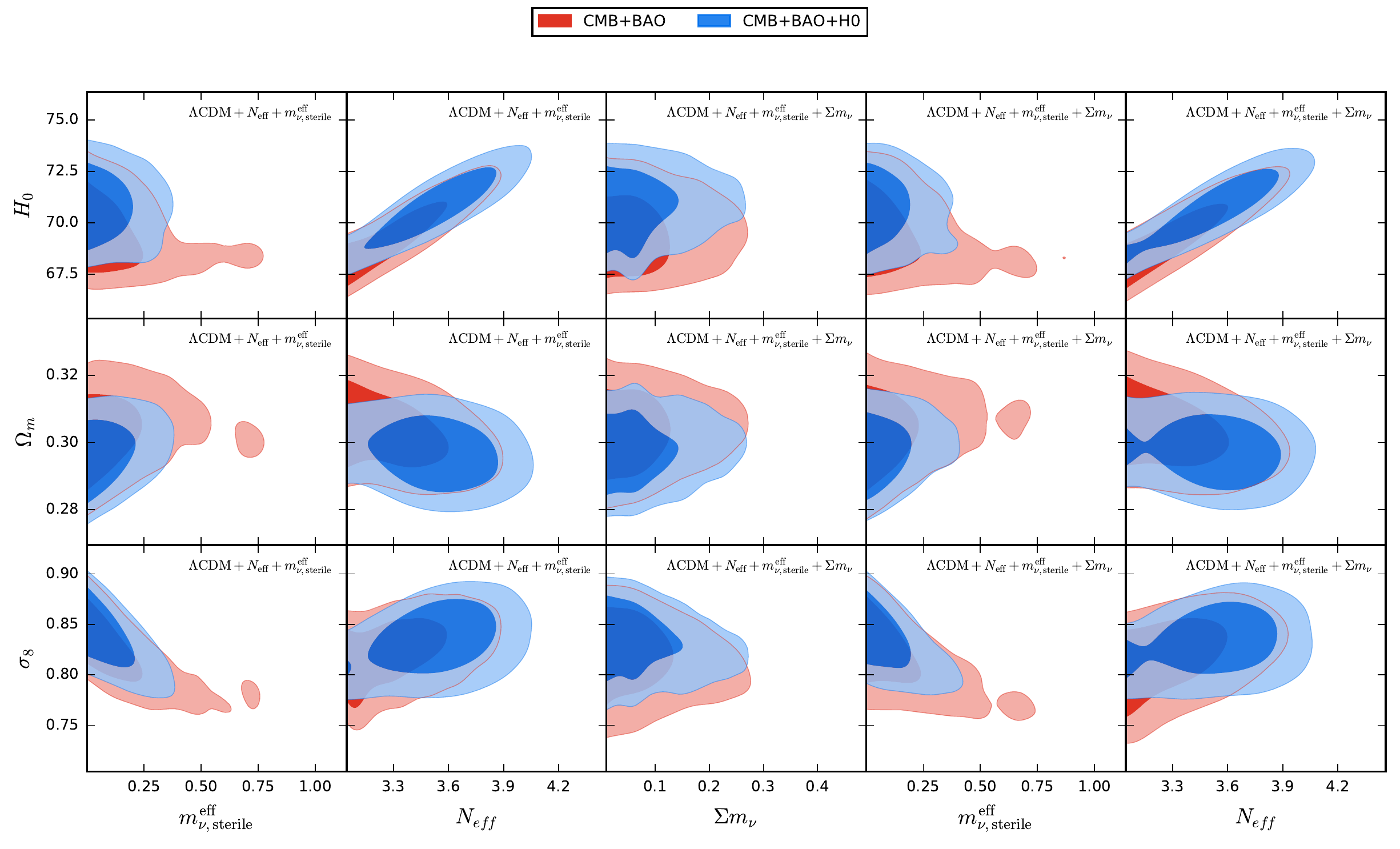} 
\caption{68 and 95 per cent confidence level constraints from \emph{Planck} CMB + BAO and \emph{Planck} CMB + BAO + $H_{0}$ on the late Universe cosmological parameters $H_{0}$, $\Omega_{m}$, $\sigma_{8}$, and the introduced parameters in the neutrino extensions of the base $\Lambda$CDM model. The comparison between marginalized contours show the parameter degeneracies and the contribution from the locally measured $H_{0}$. Here priors on parameters are chosen from base \emph{Planck} 2015 analysis.}
\label{fig:3} 

\end{figure*}

\begin{figure*}
 \begin{center}
  \subfigure [ ]{\includegraphics[width= 6.43cm,angle=360]{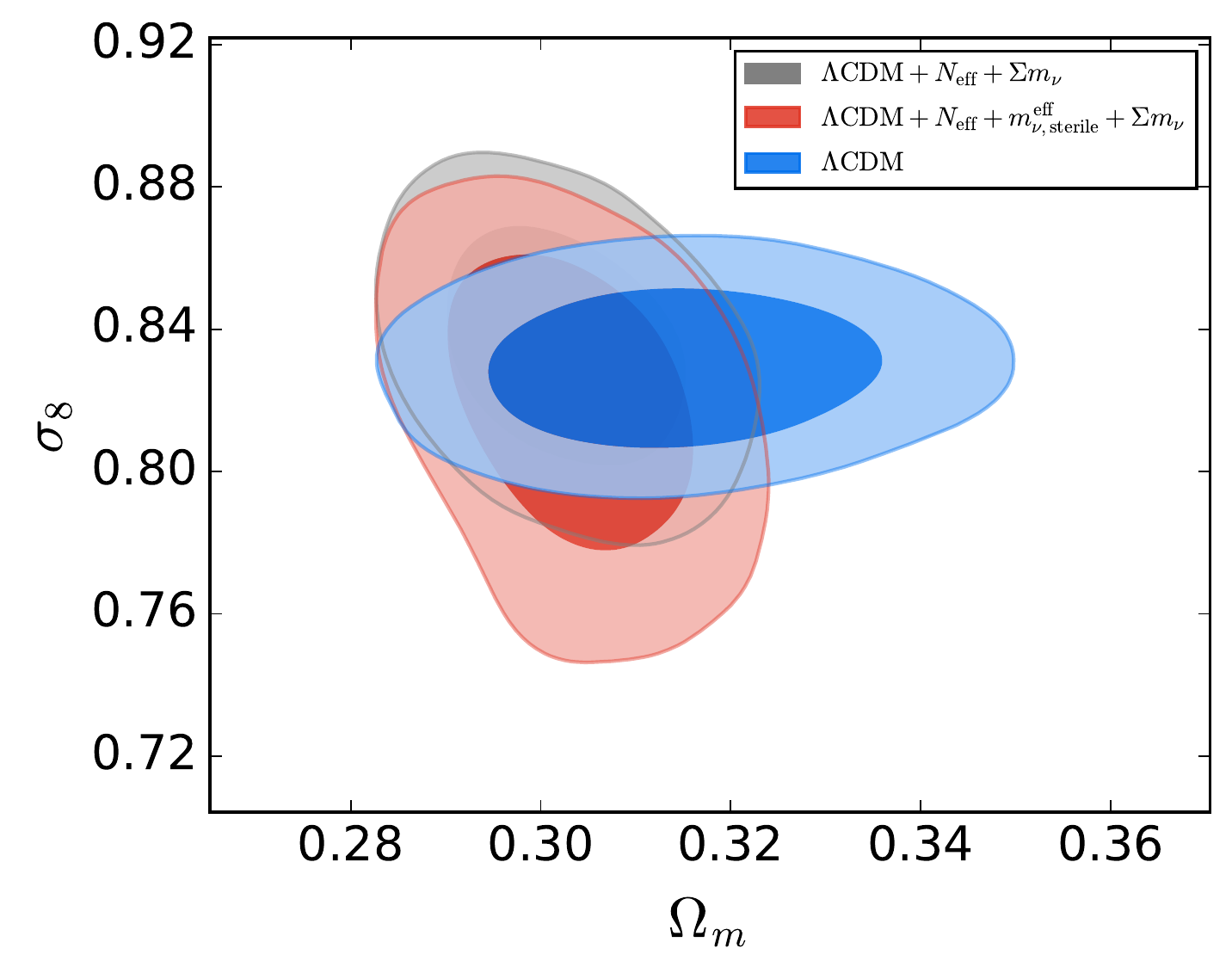}}
\subfigure [ ]{\includegraphics[width=5.66cm,angle=360]{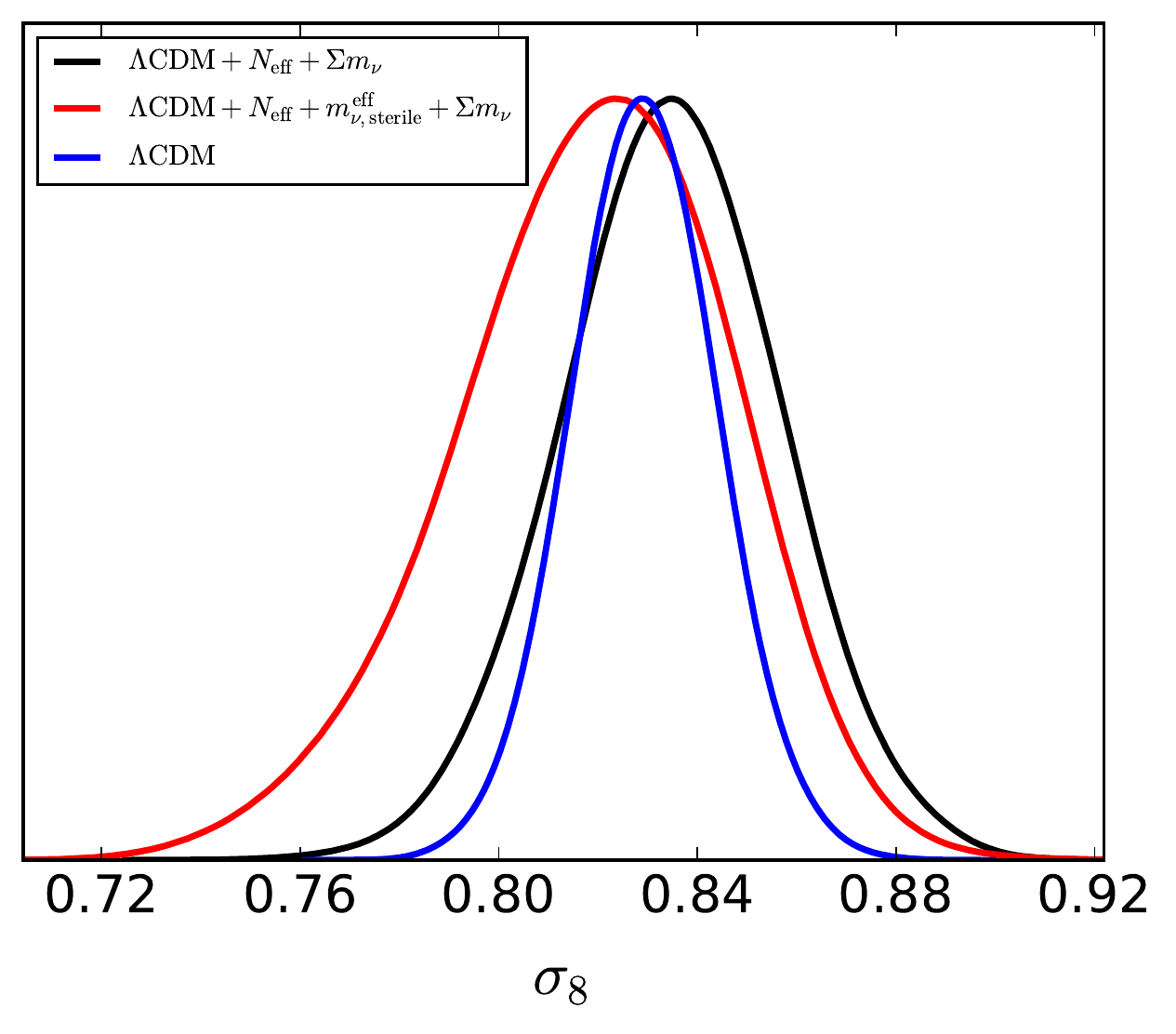}}
\caption{Panel(a): Two-dimensional marginalized posterior contours (68 and 95 per cent confidence level) of the constraint results from \emph{Planck} CMB + BAO data combination for the $\Lambda$CDM + $N_\textrm{eff}$ + $ \Sigma m_{\nu}$ model (gray contour) and the $\Lambda$CDM + $N_\textrm{eff}$ + $m_{\nu,sterile}^{eff}$ + $\Sigma m_{\nu}$ model (red contour); and from \emph{Planck} CMB data for the base $\Lambda$CDM model (blue contour), in the $\Omega_{m}$- $\sigma_{8}$ plane. It shows the significant difference on $\Omega_{m}$ constraints between the blue contour, and both, the gray and red contours, which is due to the use of the updated BAO data in this analysis. A significant difference can be seen in $\sigma_{8}$ constraints between the red contour and the gray and blue contours, which is due to introducing $m_{\nu,sterile}^{eff}$ parameter in the model (red contour). Panel(b): One-dimensional marginalized posterior distributions for $\sigma_{8}$. }
\label{fig:4}
\end{center}
\end{figure*}

Table \ref{Table:1} indicates that under the CMB data combined with updated BAO measurement in this paper, we derive larger constraint values for $\sigma_{8}$, in comparison to the results of similar analysis in baseline \emph{Planck} 2015. It also shows that the locally measured $H_{0}$ has an impact on $\sigma_{8}$, increasing its derived constraint values for our extended models, where for instance we obtain the highest value of 0.851 $\pm $ 0.019 at the 68 per cent confidence level for this parameter in the model extended with $N_\textrm{eff}$. These results also show that for the models included with parameters $m_{\nu,sterile}^{eff} $ or $\Sigma m_{\nu} $ we derive smaller constraint values for parameter $\sigma_{8}$, where a massive sterile neutrino helps to pull down the amount of constraints on this parameter more than the degenerate massive neutrinos. This effect is also discussed in earlier works such as \emph{Planck} collaboration results. In plot (a) of Fig. \ref{fig:2}, we show the impact of massive sterile neutrinos on $\sigma_{8} $ under the constraints from both data combinations in $\Lambda$CDM + $N_\textrm{eff}$ + $m_{\nu,sterile}^{eff} $. The plot indicates that $m_{\nu,sterile}^{eff} $ is negatively correlated with $\sigma_{8}$, which implies that the higher mass, for instance the massive sterile neutrino in the cosmological model, leads to a lower $\sigma_{8}$. The obtained constraint values of $\sigma_{8}$ in Table \ref{Table:1} are consistent with the observed tendency in the plot.
\par
One issue with the constraints on $m_{\nu,sterile}^{eff} $ from CMB observations, is the need for the informative physical $m_{sterile}^{thermal}$ prior.
In plot (a) of Fig. \ref{fig:2}, we assume that the observed high value of $m_{\nu,sterile}^{eff} $ is caused by this problem.
But interestingly, this plot also indicates that the additional locally measured $H_{0}$ value might resolve this problem by reducing the high values of $m_{\nu,sterile}^{eff} $.
The effect of very massive neutrinos on the CMB spectrum is identical to that of cold dark matter \citep{2016A&A...594A..13P} which cannot be distinguished.
Hence, a prior of $m_{sterile}^{thermal}$ < 10 eV on the physical mass is imposed to cut out most of the regions of parameter space in plot (b) of Fig. \ref{fig:2} where the neutrinos with high mass behave the same as dark matter; the shaded area corresponds to the excluded regions of parameter space.
In this model under data containing local $H_{0}$, the measured value for dark radiation  is pulled away from the standard value $N_\textrm{eff}$ = 3.046.
From this, we may conclude that there would be a lesser distribution probability in the high-mass tail.
Further, it suggests that there could be no need for a high physical-mass tail when constraining massive sterile neutrinos from the CMB under the locally measured $H_{0}$.\par
It is assumed that the consideration of a low redshift data combination, such as the package of SZ (the Planck Sunyaev-Zeldovich), the \emph{Planck} Lensing, WL (the weak lensing), galaxy clustering, and the local measurement of the Hubble constant $ H_{0} $, can lead to a better constraint on $\sigma_{8}$ resulting in a lower value \citep{2017MNRAS.471.4412K}. However, the results in this work show that the inclusion of an $ H_{0} $-measured at low redshift-in our data combination does not help to obtain this.

\par
Fig. \ref{fig:3} shows the plots of introduced parameters in various neutrino extensions of the base $\Lambda$CDM model against the late Universe cosmological parameters. Parameter contours show the clear degeneracies between cosmological and the introduced parameters; and the contribution from the locally measured $ H_{0} $, for instance, its significant effect on $m_{\nu,sterile}^{eff} $, decreasing high values of mass for sterile neutrino is clear.

\subsubsection{The effect on the structure growth}

The massive active neutrinos are degenerated with the cosmological parameters. They slow down the growth of the matter fluctuation \citep{2013neco.book.....L}. Here we include the massive active neutrino along with a massive sterile neutrino in the cosmological model in order to find out their contribution to $\sigma_{8}$. The plot in panel (a) of Fig. \ref{fig:4} indicates that, the model with a massive sterile neutrino can produce a lower value for parameter $ \sigma_{8}$, in comparison to the model with only massive active neutrino; Table \ref{Table:1} shows that under the constraint of CMB + BAO we obtain $ \sigma_{8}$ = $0.819^{+0.030}_{-0.025} $ and $ \sigma_{8}$ = 0.834 $ \pm $ 0.022 (68 per cent confidence level) for the two models. The plot in panel (b) of Fig. \ref{fig:4} supports the effect observed in panel (a), that the massive sterile neutrino in the model results in a constraint on $ \sigma_{8} $ that covers the lower value of $ \sigma_{8} $. Thus, under the constraint from this data combination for the model with both, $m_{\nu,sterile}^{eff}$ and $\Sigma m_{\nu}$, we produce a constraint with a lower and upper limit of $\sigma_{8} $ = $0.819^{+0.051}_{-0.060} $ at the 95 per cent confidence level. These results imply that the potential massive sterile neutrinos in cosmology could actually slow down the structure growth in the Universe more than the standard massive neutrinos. As a further result of this analysis we report the compatibility in this parameter with the constraints from \emph{Planck} 2015 CMB data at the 1 $\sigma$ level.
Although, in KiDS-450, the kilo Degree survey, the analysis of the data from the weak gravitational lensing cosmic shear power spectrum, which is based on 450 deg$^{2}$, resulted in the value $S_{8} $ $\simeq$ 0.65 $\pm$ 0.05, ($S_{8} $ $\equiv$ $ \sigma_{8} \sqrt{\Omega_{m}/0.3} $), which is in tension with the constraint of the \emph{Planck} data at the 3.2 $\sigma $ level \citep{2017MNRAS.471.4412K}. Our result is in tension with the result from the Kids-450 analysis at the $\simeq$ 3 $\sigma $ level. This shows that including parameter $m_{\nu,sterile}^{eff} $ in the model of cosmology reduces the $\sigma_{8} $ value, and moves it closer to the value from the Kids-450 for this parameter. Furthermore, our result disagrees-at the 1.9 $\sigma $ level-with the $S_{8} $ = $0.804^{+0.032}_{-0.029}$ (68 per cent confidence level) obtained from the recent cosmological analysis of cosmic shear two-point correlation functions (TPCFs) from Hyper Suprime-Cam Subaru Strategic Program (HSC SSP) first-year data, covering 136.9 deg$^{2}$ of the sky \citep{2019arXiv190606041H}. Our result also disagrees-at a 2.05 $\sigma $ level-with the value of $S_{8} $ = $0.782^{+0.027}_{-0.027}$ (68 per cent confidence level) from the results of measurements of cosmic shear in a galaxy survey from Dark Energy Survey (DES) year 1 shape catalogs, over 1321 deg$^{2}$ of the southern sky  \citep{2018PhRvD..98d3528T}
\par
The $\Omega_{m}$ constraint results in the panel (a) of the Fig. \ref{fig:4} is in discrepancy with the published constraints in baseline \emph{Planck} 2015. Here, the plot shows that there is a significant difference between the constraints for $\Lambda$CDM and $\Lambda$CDM + $N_\textrm{eff}$ + $\Sigma m_{\nu}$, whereas, the $\Omega_{m}$ constraints are very similar between these two models in \emph{Planck} results. This is due to the updated BAO data we use in this analysis which leads to the smaller constraints $\Omega_{m}$.

\section{Conclusions}

Sterile neutrinos could be a natural extension to the SM of cosmology to help better understand the properties of the Universe.
\par
As we predicted in this study, the current CMB, as well as some low redshift BAO observations, may not provide us with good information on the mass of particles such as sterile neutrinos in an extended $\Lambda$CDM model; as we only obtain an upper limit on parameter $m_{\nu,sterile}^{eff}$. In the case of three degenerate massive neutrino 
$\Sigma m_{\nu}$, we find only an upper limit on the mass of this particle too. Addition of the locally measured Hubble parameter $ H_{0} $ in the data helps to constrain the mass by only tightening the constraints on the parameters $m_{\nu,sterile}^{eff}$ and $\Sigma m_{\nu}$. Involving massive sterile neutrino in cosmology may help to slow down the rate of structure growth in the Universe, by leading to lower $\sigma_{8}$ values, where we show the compatibility in obtained $\sigma_{8}$ with the constraints from \emph{Planck} 2015 analysis. A potential massive sterile neutrino in cosmology might be helpful to resolve the tension between \emph{Planck} CMB observations and the late astrophysical measurements of $ H_{0} $ and  $\sigma_{8}$; whereas, the impact from standard massive neutrinos on these parameters is very small. We observe that the impact from locally measured $ H_{0} $ on the value of the CMB constraints of $m_{\nu,sterile}^{eff}$ might solve the issue of the need for an informative physical $m_{sterile}^{thermal}$ prior. We find that the addition of the local $ H_{0} $ in the observational data, results in increasing constraint values for $\sigma_{8}$.     

\par
A future study on the B-mode signal from CMB polarization observations might provide hints on the nature of all physical properties of the Universe that influence the formation of large structure, such as the mass of neutrinos or dark matter.

\section*{Acknowledgements}

This paper is based on the author's masters thesis, written under the supervision of Professor Antony Lewis of the Astronomy Centre in the Department of Physics and Astronomy at the University of Sussex. The author wishes to thank Professor Lewis for very helpful comments and discussions over the period of his supervision as well as the helpful hints to some questions that he provided over the preparation of this paper. Appreciation also goes to the anonymous referee for very helpful comments that improved the manuscript and its presentation. The author wishes to thank his friends, Dr Ralph Eatough from Max Planck institute for Radio Astronomy in Bonn and Dr Mark Purver, former student at the University of Manchester, for proof-reading the manuscript.

%%%%%%%%%%%%%%%%%%%%%%%%%%%%%%%%%%%%%%%%%%%%%%%%%%



\bibliographystyle{mnras}
\bibliography{mnras} % if your bibtex file is called example.bib

%\begin{thebibliography}{99}
%\end{thebibliography}

%%%%%%%%%%%%%%%%%%%%%%%%%%%%%%%%%%%%%%%%%%%%%%%%%%

%%%%%%%%%%%%%%%%% APPENDICES %%%%%%%%%%%%%%%%%%%%%

%\appendix



%%%%%%%%%%%%%%%%%%%%%%%%%%%%%%%%%%%%%%%%%%%%%%%%%%

% Don't change these lines
\bsp	% typesetting comment
\label{lastpage}
\end{document}